\documentclass[a4paper,UKenglish,cleveref, autoref, thm-restate]{lipics-v2021}
\usepackage{graphicx}
\usepackage{hyperref}
\usepackage{subfiles}
\usepackage{amsmath}
\usepackage{dsfont}
\usepackage{ifsym}
\usepackage{comment}
\usepackage[dvipsnames]{xcolor}
\usepackage{boxedminipage}
\usepackage[ruled,vlined]{algorithm2e}

\RestyleAlgo{ruled}

\SetKwComment{Comment}{/* }{ */}

\title{Exact Matching in Graphs of Bounded Independence Number} 

\author{Nicolas {El Maalouly}}{Department of Computer Science, ETH Z\"{u}rich, Switzerland }{nicolas.elmaalouly@inf.ethz.ch}{0000-0002-1037-0203
}{}
\author{Raphael {Steiner}}{Department of Computer Science, ETH Z\"{u}rich, Switzerland }{raphaelmario.steiner@inf.ethz.ch}{0000-0002-4234-6136}{supported by an ETH Zurich Postdoctoral Fellowship.}

\authorrunning{N. El Maalouly, R. Steiner} 

\Copyright{Nicolas El Maalouly, Raphael Steiner} 



\relatedversiondetails{Full Version}{https://arxiv.org/abs/2202.11988} 

\nolinenumbers 

\EventEditors{Stefan Szeider, Robert Ganian, and Alexandra Silva}
\EventNoEds{3}
\EventLongTitle{47th International Symposium on Mathematical Foundations of Computer Science (MFCS 2022)}
\EventShortTitle{MFCS 2022}
\EventAcronym{MFCS}
\EventYear{2022}
\EventDate{August 22--26, 2022}
\EventLocation{Vienna, Austria}
\EventLogo{}
\SeriesVolume{241}
\ArticleNo{65}

\begin{document}
\maketitle

\begin{abstract}

In the \emph{Exact Matching Problem} (EM), we are given a graph equipped with a fixed coloring of its edges with two colors (red and blue), as well as a positive integer $k$. The task is then to decide whether the given graph contains a perfect matching exactly $k$ of whose edges have color red. EM generalizes several important algorithmic problems such as \emph{perfect matching} and restricted minimum weight spanning tree problems.

When introducing the problem in 1982, Papadimitriou and Yannakakis conjectured EM to be \textbf{NP}-complete. Later however, Mulmuley et al.~presented a randomized polynomial time algorithm for EM, which puts EM in \textbf{RP}. Given that to decide whether or not \textbf{RP}$=$\textbf{P} represents a big open challenge in complexity theory, this makes it unlikely for EM to be \textbf{NP}-complete, and in fact indicates the possibility of a \emph{deterministic} polynomial time algorithm. EM remains one of the few natural combinatorial problems in \textbf{RP} which are not known to be contained in \textbf{P}, making it an interesting instance for testing the hypothesis \textbf{RP}$=$\textbf{P}. 

Despite EM being quite well-known, attempts to devise deterministic polynomial algorithms have remained illusive during the last 40 years and progress has been lacking even for very restrictive classes of input graphs. In this paper we push the frontier of positive results forward by proving that EM can be solved in deterministic polynomial time for input graphs of bounded independence number, and for bipartite input graphs of bounded bipartite independence number. This generalizes previous positive results for complete (bipartite) graphs which were the only known results for EM on dense graphs.

\end{abstract}

\section{Introduction}

The problem of deciding whether a given graph contains a perfect matching, as well as the related problem of computing a maximum (minimum) weight perfect matching in a given graph are amongst the foundational problems in algorithmic graph theory and beyond, and the fact that they can be solved in polynomial time~\cite{edmonds} is an integral part of many efficient algorithms in theoretical computer science.

In 1982, Papadimitriou and Yannakakis~\cite{papadimitriou1982complexity} studied a decision problem related to perfect matchings in edge-colored graphs as follows: Given as input a graph $G$ whose edges come with a given fixed two-edge coloring (say, with colors red and blue), then the task is to decide whether for a given integer $k$ there exists a perfect matching $M$ of $G$ such that exactly $k$ of the edges in $M$ are red. Clearly, in the special case when all edges are colored red and $k=\frac{n}{2}$, this problem is simply to decide whether there exists a perfect matching in a given graph. For a heterogeneous coloring of the edges, however, the difficulty of the problem seems to change quite dramatically (see below). 

The original motivation of Papadimitriou and Yannakakis~\cite{papadimitriou1982complexity} to study the above problem, which from now on will be called \emph{Exact Matching} and abbreviated by EM, was their investigation of \emph{restricted minimum weight spanning tree problems}. In the usual minimum weight spanning tree problem, we are given a graph with non-negative edge-weights and seek to find a spanning tree minimizing the total edge-weight, and this is well-known to be solvable in polynomial time using for instance Kruskal's algorithm~\cite{kruskal}. Papadimitriou and Yannakakis considered what happens if we restrict the shape of the spanning trees allowed in the output, and obtained several results. For instance, the problem is easily seen to be \textbf{NP}-hard if the considered spanning trees are constrained to be paths, by a reduction from the Hamiltonian Path problem, but it is polynomial-time solvable if the tree shapes are restricted to stars or $2$-stars. While for many classes of trees, Papadimitriou and Yannakakis~\cite{papadimitriou1982complexity} classified the complexity of the above problem, some cases remained unsettled. In particular, they proved that the restricted minimum weight spanning tree problem for so-called \emph{double $2$-stars} is equivalent to EM, and left it as an open problem to decide its computational complexity. In fact, they stated the conjecture that EM is \textbf{NP}-complete. Up until today, neither has this conjecture been confirmed, nor is it known whether EM can be solved in polynomial time by a deterministic algorithm. Yet, there have been some interesting results and developments regarding the problem in the past, which we summarize in the following.

Only few years after the introduction of the problem, in a breakthrough result Mulmuley, Vazirani and Vazirani~\cite{mulmuley1987matching} developed their so-called \emph{isolation lemma}, and demonstrated its power by using it to prove that EM can be solved by a randomized polynomial time algorithm, i.e. it is contained in \textbf{RP}. This makes it unlikely to be \textbf{NP}-hard. In fact, deciding whether \textbf{RP}$=$\textbf{P} remains one of the big challenges in complexity theory.
This means that problems such as EM, for which we know containment in \textbf{RP} but are not aware of deterministic polynomial time algorithms, are interesting candidates for testing the hypothesis \textbf{RP}$=$\textbf{P}. Indeed, due to this, EM is cited in several papers as an open problem. This includes recent breakthrough papers such as the seminal work on the parallel computation complexity of the matching problem~\cite{svensson2017matching}, works on planarizing gadgets for perfect matchings~\cite{gurjar2012planarizing}, works on more general constrained matching problems~\cite{berger2011budgeted,mastrolilli2012constrained,mastrolilli2014bi,stamoulis2014approximation} and on multicriteria optimization problems~\cite{grandoni2010optimization} among others. Even though EM has caught the attention of many researchers from different areas, there seems to be a substantial lack of progress on the problem even when restricted to very special subclasses of input graphs as we will see next. This highlights the surprising difficulty of the problem given how simple it may seem at first glance.


\subparagraph{Previous results for EM on restricted classes of graphs.}  
It may surprise some readers that EM is even non-trivial if the input graphs are complete or complete bipartite graphs: In fact, at least four different articles have appeared on resolving these two special cases of EM~\cite{karzanov1987maximum,yi2002matchings,geerdes,gurjar2017exact}, which are now known to be solvable in deterministic polynomial time. Another positive result follows from the existence of Pfaffian orientations and their analogues on planar graphs and $K_{3,3}$-minor free graphs~\cite{yuster2012almost}, EM is solvable in polynomial time on these classes via a derandomization of the techniques used in~\cite{mulmuley1987matching}. Considering a generalization of Pfaffian orientations, it was further proved in~\cite{genus} that EM can be solved in polynomial time for graphs embeddable on a surface of bounded genus. Finally, from the well-known meta-theorem of Courcelle~\cite{courcelle}, one easily obtains that EM can be efficiently solved on classes of bounded tree-width. 




\subparagraph{Our contribution.} In this paper, we generalize the known positive results for EM on very dense graphs such as complete and complete bipartite graphs to graphs of independence number at most $\alpha$ and to bipartite graphs of bipartite independence number at most $\beta$, for all fixed integers $\alpha, \beta \ge 1$. The \emph{independence number} of a graph $G$ is defined as the largest number $\alpha$ such that $G$ contains an \emph{independent set} of size $\alpha$. The \emph{bipartite independence number} of a bipartite graph $G$ equipped with a bipartition of its vertices is defined as the largest number $\beta$ such that $G$ contains a \emph{balanced independent set} of size $2\beta$, i.e., an independent set using exactly $\beta$ vertices from both color classes.

\begin{theorem} \label{th:XPalpha}
There is a deterministic algorithm for EM on graphs of independence number $\alpha$ running in time $n^{O(f(\alpha))}$, for $f(\alpha)=2^{O(\alpha)}$.
\end{theorem}

\begin{theorem}\label{th:XPbeta}
There is a deterministic algorithm for EM on bipartite graphs of bipartite independence number $\beta$ running in time $n^{O(f(\beta))}$, for $f(\beta)=2^{O(\beta)}$.
\end{theorem}

The special cases $\alpha=1$ and $\beta=1$ of the above results correspond exactly to the previously studied cases of complete and complete bipartite graphs. We emphasize that even though bounding the independence number might seem like a big restriction on the input graphs, already for $\alpha=2, \beta=2$ our results cover rich and complicated classes of graphs, for instance every complement of a triangle-free graph belongs to the class of independence number at most $2$, and every bipartite complement of a $C_4$-free bipartite graph belongs to the class of bipartite independence number at most $2$. 

Another interesting observation in support of the above is the following: So far, for all classes of graphs on which EM was known to be solvable in polynomial time (including planar graphs, $K_{3,3}$-minor-free graphs, graphs of bounded genus, complete and complete bipartite graphs), the number of perfect matchings was also known to be countable in polynomial time (cf.~\cite{kasteleyn,little,genus,yuster2012almost}), and one may wonder about whether tractability of EM aligns with the tractability of corresponding counting problems for perfect matchings. However, even for graphs of independence number $2$ we are not aware that polynomial schemes for counting perfect matchings exist, and in fact conjecture that this problem is computationally hard, therefore putting our result into nice contrast with previous positive results on EM.

\begin{conjecture}\label{conj}
The problem of counting perfect matchings in input graphs of independence number $2$ is \#\textbf{P}-complete.
\end{conjecture}

\subparagraph{Organization of the paper.}

The remainder of this paper is organized as follows: 
In \Cref{sec:Prel} we present the basic definitions and conventions we use throughout the paper.
In \Cref{sec:XPalpha} we prove \Cref{th:XPalpha}, i.e., showing the existence of an XP algorithm parameterized by the independence number of the graph.
In \Cref{sec:XPbeta} we consider the bipartite graphs case and prove \Cref{th:XPbeta}.
In \Cref{sec:distD} we discuss distance-$d$ independence number parameterizations and in \Cref{sec:conc} we conclude the paper and provide some open problems.

\section{Preliminaries}\label{sec:Prel}



Due to space restrictions, proofs of statements marked $\star$ have been deferred to
the appendix.
All graphs considered are simple. For a graph $G= (V,E)$ we let $n= |V(G)|$, i.e. the number of vertices in $G$. Given an instance of EM and a perfect matching\footnote{A perfect matching of a graph is a matching (i.e., an independent edge set) in which every vertex of the graph is incident to exactly one edge of the matching.} (abbreviated PM) $M$, we define edge weights as follows: blue edges get weight 0, matching red edges get weight $-1$ and non-matching red edges get weight $+1$.
For $G'$ a subgraph\footnote{Note that the subgraph can also be a set of edges or cycles.} of $G$, we define $R(G')$ (resp. $B(G')$) to be the set of red (resp. blue) edges in $G'$, $r(G') := |R(G')|$ and $w_M(G')$ to be the sum of the weights of edges in $G'$. For ease of notation, we will use $w(G')$ for $w_M(G')$ and will make the matching explicit whenever it is not $M$.

Whenever we consider a set of cycles or paths, it is always assumed that they are vertex disjoint and alternating with respect to the current matching $M$ (unless specified otherwise).
Define an $x$-path to be an alternating path of weight $x$.
Undirected cycles are considered to have an arbitrary orientation. For a cycle $C$ and $u,v \in C$, $C[u,v]$ is defined as the path from $u$ to $v$ along $C$ (in the fixed but arbitrarily chosen orientation). For simplicity, a cycle is also considered to be a path i.e. a closed path (its starting vertex is chosen arbitrarily).
$Ram(r,s)$ refers to the Ramsey number, i.e. every graph on $Ram(r, s)$ vertices contains either a clique of size $r$ or an independent set of size $s$. For simplicity we will use the following upper bound $Ram(s+1, s+1) < 4^s$ \cite{do2019party}.
For two sets of edges $M$ and $M'$, $M\Delta M'$ refers to the symmetric difference between the two sets (i.e. the edges that appear in exactly one of the two sets). Note that if $M$ and $M'$ are two PMs, then $M\Delta M'$ forms a set of cycles (each alternating with respect to both matchings) and will be use as such. Also note that with the above defined edge weights we have $r(M') = r(M) + w(M\Delta M')$.

\section{Bounded Independence Number Graphs}\label{sec:XPalpha}

The algorithm relies on a 2 phase process. The first phase is an algorithm that outputs  a PM $M$ with $|k- r(M)|$ bounded (by a function of $\alpha$), i.e. with a number of red edges that only differs from $k$ by a function of $\alpha$. This algorithm is also of independent interest since it provides a solution that is close to optimal (for small independence number) while its running time is polynomial and independent of the independence number. 

\begin{theorem} \label{th:falphaclose}
Given a `Yes' instance of EM, there exists a deterministic polynomial time algorithm that outputs a PM $M$ with  $k-2\cdot 4^{\alpha} \leq r(M) \leq k$.
\end{theorem}

\begin{remark*}
Note that a standalone proof of \Cref{th:falphaclose} can be made quite simple but would require additional notions and definitions. 
Our main focus however, is on the proof of \Cref{th:XPalpha}, so the proof structure is tailored towards that end and the proof of \Cref{th:falphaclose} will come as result along the way.
\end{remark*}

The second phase is an algorithm that outputs a solution matching with a running time that depends on the size of the smallest color class in a symmetric difference between a given matching and a solution matching. It is also of independent interest as it can be more generally useful for the study of other parameterizations of EM as well as other matching problems with color constraints.
 
\begin{proposition} \label{prop:smallsetedges}
Let $M$ and $M'$ be two PMs in $G$ s.t. $|B(M \Delta M')| \leq L$ or $|R(M \Delta M')| \leq L$. Then there exists a deterministic algorithm running in time $n^{O(L)}$ such that given $M$ it outputs a PM $M''$ with $r(M'') = r(M')$. 
\end{proposition}

\begin{proof}
Suppose w.l.o.g. $|R(M \Delta M')| \leq L$ (the other case is similar by swapping the colors). Guess $R(M \Delta M')$ in time $n^{O(L)}$ by trying all possibilities (the rest of the algorithm should succeed for at least one such possibility). Compute $R(M') = R(M) \Delta R(M \Delta M')$. Remove the edges $R(M')$ and their endpoints from the graph as well as all remaining red edges. Compute a PM $M_1$ on the rest of the graph (such a PM must exist since $M' \backslash R(M')$ is one such example) and let $M'' := M_1 \cup R(M')$. Observe that $M''$ is a PM with $r(M'') = |R(M'')| = |R(M')| = r(M')$.
\end{proof}


For this phase to run in polynomial time for bounded independence number, we need to show that there exists a PM $M^*$ with exactly $k$ red edges, where $M \Delta M^*$ ($M$ being the PM we get after the first phase) has a bounded (by a function of $\alpha$) number of edges of some color class. 
The main technical challenge is to show that for this to be the case it is sufficient to have $|k- r(M)|$ bounded (which is guaranteed by the first phase). The rest of this section is devoted to this proof. Along the way we will also prove \Cref{th:falphaclose}. Before going into the technical details, we give a quick overview.
 
\subsection{Proof Overview}

In order to apply \Cref{prop:smallsetedges}, we will consider the solution matching $M^*$ which minimizes the number of edges in $M \Delta M^*$ ($M$ being the PM we get after the first phase) and aim to show that it contains a bounded number of edges of some color class. Towards this end, we want to show that if the set of alternating cycles $M \Delta M^*$ contains a large number of edges from both color classes, then there be must another set of alternating cycles $\mathcal{C}$ with the same total weight as $M \Delta M^*$, but containing strictly less edges. This contradicts the minimality of $M \Delta M^*$ since $M \Delta \mathcal{C}$ is also a solution matching with $|E(M \Delta (M \Delta \mathcal{C}))| = |E(\mathcal{C})| < |E(M \Delta M^*)|$. In other words, we want to show that unless one color class in $M \Delta M^*$ is bounded, we can reduce the size of one or more of the cycles in $M \Delta M^*$ while keeping the total weight unchanged.

\subparagraph{Skips.}

\begin{figure}[ht]
    \centering
    \includegraphics[scale=0.75]{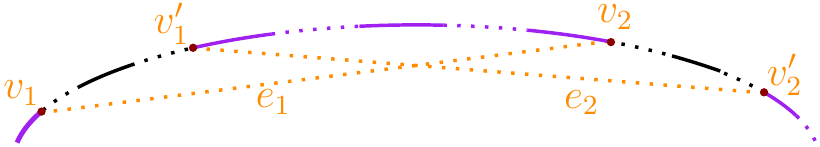}
    \caption{A skip formed by two non-matching edges $e_1$ and $e_2$ (in orange). Matching edges are represented by full lines and non-matching edges by dotted lines. The paths removed by the skip are depicted in black while the rest of the alternating cycle is depicted in purple.}
    \label{fig:skip}
\end{figure}

The main tool we use to show the existence of smaller alternating cycles is something we call a skip (see \Cref{fig:skip}). At a high level, a skip is simply a pair of edges that creates a new alternating cycle $C'$ by replacing two paths of an alternating cycle $C$. If those paths have total length more than 2 then $|C'| < |C|$. This means that if a solution matching $M^{**}$ exists, such that $M \Delta M^{**}$ is the same as $M \Delta M^*$ but with $C$ replaced by $C'$, it would contradict the minimality of $M \Delta M^*$. For $M^{**}$ to be a solution matching, we also need $w(C) = w(C')$ so that $M^{**}$ also has $k$ red edges. For this reason we look for skips that do not change the total weight (we call them 0 skips). It can happen however, that even though no 0 skip exists, a collection of skips exists, that can be used independently, and their total weight change is zero (we call them 0 skip sets). Also observe that these skips can come from different cycles of $M \Delta M^*$ and still be used to reduce its total number of edges (i.e. we can modify multiple cycles in $M \Delta M^*$ simultaneously to preserve the total weight change). 
So by taking $M \Delta M^*$ to be minimal (in terms of total number of edges), we are guaranteed that no such skip sets can exist.

\subparagraph{Skips from Paths.}
To show the existence of skips (which will lead to the desired contradiction), we rely on Ramsey theory to show that if we take a large enough (with respect to $\alpha$) collection of disjoint paths on an alternating cycle, starting and ending with non-matching edges, then they must form skips. Now if these paths have certain desired weights, then we could make sure that we get a 0 skip set as desired.

\subparagraph{Paths from Edge Pairs.}
To prove the existence of paths of desired weight, we analyze the cycles in $M \Delta M^*$ by looking at their edge pairs, i.e. pairs of consecutive matching and non-matching edges. These edge pairs can have 3 configurations from which we can extract the paths. (1) Consecutive same sign pairs (sign here refers to the weight of the pair), (2) consecutive different sign pairs and (3) consecutive 0 pairs. We show that we can extract paths of the desired properties from all of these configurations, and the types of skips we get is dependent on the weights of the cycles and the sizes of their color classes.

\subparagraph{Bounding the Cycle Weights.}
Next, we show that if $M \Delta M^*$ is minimal, all of its cycles have bounded weight. This is mainly achieved by showing that cycles of large weight must have skips that reduce their weight. This changes the total weight of $M \Delta M^*$ however, and must be compensated for either by skips on a cycle of opposite sign weight, or by removing some of the cycles in $M \Delta M^*$.

\subparagraph{Bounding one color class.}
After bounding the weights of the cycles in $M \Delta M^*$ (by a function of $\alpha$), we will also bound their number given that their total weight is bounded. With these properties (bounded cycle weights and number of cycles), we can show that if $M \Delta M^*$ has enough edges from both colors, then at least one of its cycles contains enough positive skips and one of its cycles contains enough negative skips, together forming a 0 skip set, i.e. $M \Delta M^*$ is not minimal. So choosing $M \Delta M^*$ minimal implies a bound on the size of one of its color classes.

\subsection{Detailed Proof}

\subparagraph{Skips.} We start by formally defining a skip and its properties.

\begin{definition} \label{def:skip}
Let $C$ be an alternating cycle. A skip $S$ is a set of two non-matching edges $e_1 := (v_1,v_2)$ and $e_2 := (v_1',v_2')$ with $e_1, e_2 \notin C$ and $v_1,v_1',v_2,v_2' \in C$ (appearing in this order along $C$) s.t. $C' = e_1 \cup e_2 \cup C \setminus (C[v_1,v_1'] \cup C[v_2,v_2'])$ is an alternating cycle, $|C| - |C'| > 0$ and $|w(S)| \leq 4$ where $w(S):= w(C') - w(C)$ is called the weight of the skip. 
\end{definition} 
Note that we require a skip to have weight at most 4. This is mainly to simplify the analysis since it is enough to only consider such skips.

If $P \subseteq C$ is a path and $C[v_1,v_2'] \subseteq P$, then we say that $P$ contains the skip $S$.
We say using $S$ to mean replacing $C$ by $C'$. If $C \in M\Delta M'$ for some PM $M'$, then by using $S$ we also modify $M'$ accordingly (i.e. s.t. $M \Delta M'$ now contains $C'$ instead of $C$). Observe that a positive skip (where positive refers to the weight of the skip) increases the cycle weight, a negative skip decreases it and a 0 skip does not change the cycle weight. Using a skip always results in a cycle of smaller cardinality.
Two skips $\{(v_1,v_2),(v_1',v_2')\}$ and $\{(u_1,u_2),(u_1',u_2')\}$ are called disjoint if they are contained in disjoint paths along the cycle. Note that two disjoint skips can be used independently. 

\begin{definition}
Let $\mathcal{C}$ be a set of alternating cycles. A 0 skip set is a set of disjoint skips on cycles of $\mathcal{C}$ s.t. the total weight of the skips is 0.
\end{definition}

Observe that finding a skip with some desired properties can be done in polynomial time by trying all possible combinations of 2 edges, every time checking if the edges form a skip with the desired properties (i.e. checking if the resulting cycle $C'$ is alternating, has strictly less edges then $C$ and the weight change is as desired, which can all be done in polynomial time).

\subparagraph{Skips from Paths.}

Next, we show that if a cycle contains a lot of disjoint paths then it must contain a skip that replaces 2 of these paths by its 2 edges.
\begin{figure}[ht]
    \centering
    \includegraphics[scale=0.75]{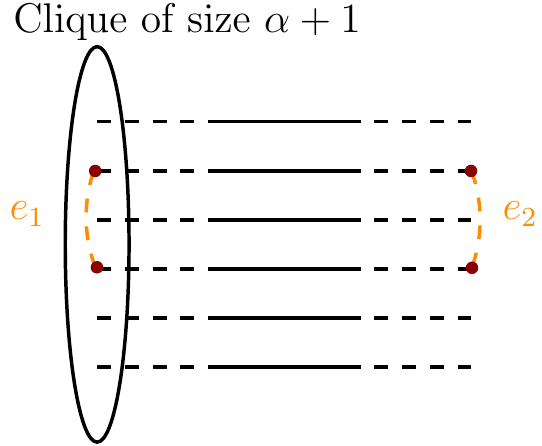}
    \caption{A subset of $\mathcal{P}$ of size $\alpha + 1$ whose starting vertices form a clique.}
    \label{fig:Ramsey}
\end{figure}
\begin{lemma}\label{lem:skipfrompaths}
Let $P$ be an alternating path containing a set $\mathcal{P}$ of disjoint paths, each of length at least $3$ and starting and ending at non-matching edges, of size $|\mathcal{P}| \geq 4^{\alpha}$. Then $P$ contains a skip.
If all paths in $\mathcal{P}$ have the same weight $x$, then if $x$ is one of the following values, we get the following types of skips:
\begin{itemize}
    \item $x=2$: negative skip.
    \item $x=1$: negative or 0 skip.
    \item $x=0$: positive or 0 skip.
    \item $x=-1$: positive skip.
\end{itemize}
\end{lemma}

\begin{proof}
The set of starting vertices of the paths in $\mathcal{P}$ must contain a clique $Q$ of size $\alpha + 1$ since $|\mathcal{P}|>Ram(\alpha+1, \alpha+1)$ (and the independence number of the graph is $\alpha$). Let $\mathcal{P}'$ be the set of paths from $\mathcal{P}$ starting with vertices in $Q$ and $Q'$ their set of ending vertices. Since $|Q'| = \alpha + 1$, there must be an edge connecting two of its vertices, call it $e_2$. Let $P_1$ and $ P_2$ be the two paths in $\mathcal{P}'$ connected by $e_2$. Let $e_1$ be the edge connecting the starting vertices of $P_1$ and $P_2$ (which must exist since $Q$ is a clique). Note that $e_1$ and $e_2$ must be non-matching edges since they are chords of the alternating cycle $C$ so their endpoints are matched to edges of $C$. Now observe that $e_1$ and $e_2$ form a skip $S$ (see \Cref{fig:Ramsey}) and $w(S) = w(e_1) + w(e_2) - w(P_1) - w(P_2)$. Finally, suppose $P_1$ and $P_2$ have weight $x$ and note that $w(e_1),w(e_2) \in \{0,1\}$ since they are non-matching edges. We get $-2x \leq w(S) \leq 2-2x$ thus proving the lemma.
\end{proof}

The above lemma only shows the existence of a skip of a certain sign, and does not guarantee the existence of 0 skips, i.e. skips that do not change the cycle weight. The next lemma shows that if there are enough disjoint positive and negative skips we can still obtain a 0 skip set (i.e. we can still reduce the cardinality of $M \Delta M'$ without changing its weight).

\begin{lemma}[$\star$]\label{lem:0skipset}
Let $\mathcal{S}$ be a collection of disjoint skips. If $\mathcal{S}$ contains at least 4 positive skips and at least 4 negative skips (all mutually disjoint), then $\mathcal{S}$ must contain a 0 skip set. 
\end{lemma}

\subparagraph{Edge Pairs.}
\begin{figure}[ht]
    \centering
    \includegraphics{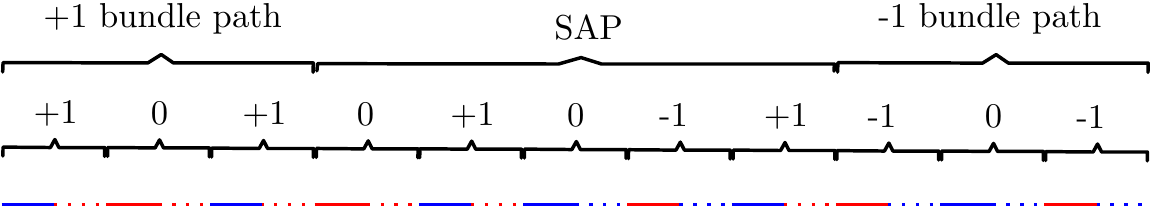}
    \caption{An example of an alternating path containing a $+1$ bundle, a $-1$ bundle and an SAP. Matching edges are represented by full lines and non-matching edges by dotted lines. The colors of the edges correspond to their color in the graph.}
    \label{fig:bundlesSAP}
\end{figure}
For a given alternating cycle, our goal is to find paths of some desired weight in order to apply \Cref{lem:skipfrompaths}. To make finding these paths easier, we look at pairs of edges. Each pair consists of two consecutive edges (the first a matching-edge and the second a non-matching edge). We label the pairs according to their weight (see in \Cref{fig:bundlesSAP} the label above each pair of edges). 
\begin{definition}
A $+1$ pair (resp. $-1$ pair and $0$ pair) is a pair of consecutive edges (the first a matching-edge and the second a non-matching edge) along an alternating cycle such that their weight sums to $1$ (resp. $-1$ and $0$).
\end{definition}

Two $+1$ (resp. $-1$) pairs are called consecutive if there is an alternating path between them on the cycle which only contains 0 pairs.
\begin{definition}
A $+1$ (resp. $-1$) bundle is a pair of edge-disjoint consecutive $+1$ (resp. $-1$) pairs. The path starting at the first pair and ending at the second one (including both pairs) is referred to as the bundle path (see \Cref{fig:bundlesSAP} for an example of such bundles).
\end{definition}
Note that a $+1$ (resp. $-1$) bundle path has weight $+2$ (resp. $-2$). Two bundles are called disjoint if their bundle paths are edge disjoint. 

\begin{definition}
A Sign Alternating Path (SAP) is an alternating path $P$ formed by edge pairs, such that it does not contain any bundles (see \Cref{fig:bundlesSAP} for an example of such path). 
\end{definition}
Note that for an SAP $P$, $|w(P)| \leq 1$.

\subparagraph{Paths from Edge Pairs.}

Our goal is to bound the number of edges from some color class in $M \Delta M^*$, when the latter is chosen to contain a minimum number of edges. To this end, we aim to show that a large number of edges of both color classes implies the existence of a 0 skip set (which would contradict the minimality of $M \Delta M^*$). By \Cref{lem:0skipset} it suffices to show the existence of many positive and negative skips which in turn can be a result of many paths of certain weight (by \Cref{lem:skipfrompaths}).

In the next two lemmas, we first show that a large number of edges of some color class implies the existence of either many bundles, a long SAP or many 0-paths starting with an edge of that color class. Then we show that all of these structures result in paths of the desired weights.

\begin{lemma}[$\star$]\label{lem:manyzeros}
Let $P$ be an alternating path containing at least $10t^3$ blue (resp. red) edges. Then one of the following properties must hold:
\begin{itemize}
    \item[(a)] $P$ contains at least $t$ disjoint bundles.
    \item[(b)] $P$ contains an SAP
    with at least $t$ non-zero pairs.
    \item[(c)] $P$ contains at least $t$ edge-disjoint 0-paths of length at least 4 starting with a blue (resp. red) matching edge.
\end{itemize}
\end{lemma}

\begin{lemma}[$\star$]\label{lem:manypaths}
A path $P$, satisfying one of the following properties, must contain $t$ disjoint paths each of length at least $3$, starting and ending with non-matching edges and having specific weights that depend on the satisfied property:
\begin{itemize}
    \item[(a)] $P$ contains $t$ disjoint $+1$ bundles: paths of weight $+2$.
    \item[(b)] $P$ contains $t$ disjoint $-1$ bundles: paths of weight $-1$.
    \item[(c)] $P$ contains $t$ edge-disjoint 0-paths of length at least 4 starting with a red matching edge: paths of weight $+1$.
    \item[(d)] $P$ contains $t$ edge-disjoint 0-paths of length at least 4 starting with a blue matching edge: paths of weight $0$.
    \item[(e)] $P$ contains an SAP with at least $2t+1$ non-zero pairs: paths of weight $+1$.
    \item[(f)] $P$ contains an SAP with at least $2t+1$ non-zero pairs: paths of weight $0$.
\end{itemize}

\end{lemma}

While the above lemmas would be enough to show the existence of many skips whenever $M \Delta M^*$ contains many edges from both color classes, these skips can still be of the same sign (e.g. all positive) which is not enough to use \Cref{lem:0skipset}. We will later show that this cannot happen if the cycles in $M \Delta M^*$ have bounded weight.

\subparagraph{Bounding Cycle Weights.}

In this part, we will deal with cycles of unbounded weight. We start by showing that a large cycle weight implies the existence of many skips that can be used to reduce it.

\begin{lemma}[$\star$]\label{lem:bigcycleskip}
Let $P$ be an alternating path with $w(P) \geq 2t\cdot 4^{\alpha}$ (resp. $w(P) \leq -2t\cdot 4^{\alpha}$), then $P$ contains at least $t$ disjoint negative (resp. positive) skips. 
\end{lemma}

The above lemma also allows for simple proof of \Cref{th:falphaclose}.

\begin{proof} [Proof of \Cref{th:falphaclose}]
Let $M_1$ be a PM containing a minimum number of red edges and $M_2$ a PM with a maximum number of red edges (should be at least $k$). Note that $M_1$ (resp. $M_2$) can be computed in polynomial time by simply using a maximum weight perfect matching algorithm with $-1$ (resp. $+1$) weights assigned to red edges and $0$ weights assigned to blue edges. 

Now as long as $r(M_1) \leq k-2\cdot 4^{\alpha}$ and $r(M_2) > k$ we will apply the following procedure (otherwise we output $M := M_1$):
Let $C \in M_1 \Delta M_2$ with $w(C) > 0$ (such a cycle must exist since $r(M_1) < r(M_2)$).
If $w(C) \leq 2\cdot 4^{\alpha}$ then we replace $M_1$ by $M_1 \Delta C$ and iterate (note that $r(M_1) < r(M_1 \Delta C) \leq k$).
Otherwise, by \Cref{lem:bigcycleskip}, $C$ contains a negative skip. We find it in polynomial time and use it to reduce the cycle weight, and iterate the whole procedure (note that $r(M_2)$ decreases). 
If at any point $r(M_2)$ drops below $k$, we simply output $M := M_2$.
In all cases $w(M_1 \Delta M_2)$ decreases after every iteration. So there can be at most $n$ iterations (since the PMs have at most $n/2$ edges each, so $w(M_1 \Delta M_2) \leq n$ and we only iterate as long as it is bigger than $0$), each running in polynomial time.

\end{proof} 

\begin{remark*}
Note that the proof only relies on \Cref{lem:bigcycleskip} which in turn only relies on \Cref{lem:skipfrompaths} and the part of \Cref{lem:manypaths} that deals with bundles. Most of the previously defined notions are not needed for this standalone result.
\end{remark*}

From \Cref{lem:bigcycleskip} we get that if $M \Delta M^*$ contains both a positive cycle of unbounded weight and a negative cycle of unbounded weight, we can find a 0 skip set using \Cref{lem:0skipset}. It could be the case however, that we have only one of the two, say a positive cycle of unbounded weight (with respect to $\alpha$), and many negative weight cycles (which would be required if $|w(M \Delta M^*)|$ is bounded, which is guaranteed by the first phase of the algorithm). In this case we can get many negative skips from the positive weight cycle of unbounded weight but we are not guaranteed to find positive skips, so we need another way to compensate for the total weight change. Notice that this can be achieved by removing negative cycles from $M \Delta M^*$. So we will combine the use of negative skips with the removal of some of the negative cycles in order to get a zero total weight change. We call this a 0 skip-cycle set.

\begin{definition}
Let $\mathcal{C}$ be a set of alternating cycles. A 0 skip-cycle set is a set of disjoint skips on cycles of $\mathcal{C}$ and/or cycles from $\mathcal{C}$, s.t. the total weight of the skips minus the total weight of the cycles is 0.
\end{definition}
We say that we use a skip-cycle set $\mathcal{S}$ to mean that we use all skips in $\mathcal{S}$ and remove all cycles in $\mathcal{S}$ from $\mathcal{C}$.
Note that a 0 skip set is a 0 skip-cycle set. Also a cycle $C \in M \Delta M^*$ with $w(C) = 0$ is a 0 skip-cycle set.
The following lemma shows that if a set of alternating cycles has bounded weight but one of its cycles has an unbounded weight then it must contain a 0 skip-cycle set.

\begin{lemma}[$\star$]\label{lem:0skipSet2}
Let $t \geq 8\cdot4^{\alpha}$ and $t' = 4t^2$.
Let $\mathcal{C}$ be a set of alternating cycles and $C \in \mathcal{C}$ s.t. $|w(\mathcal{C})| \leq t'$ and $|w(C)| \geq 2t'$, then $\mathcal{C}$ contains a 0 skip-cycle set. 

\end{lemma}

\subparagraph{Bounding one color class}

So far we have shown that we can bound the weight of the cycles in $M \Delta M^*$ (if $M \Delta M^*$ is minimal).
What we want to show next is that if $M \Delta M^*$ contains many blue (resp. red) edges and all of its cycles have bounded weight, then it also contains many positive (resp. negative) skips. This way we show that having many edges of both color classes results in a 0 skip set. 

First we deal with the case when the number of cycles in $M \Delta M^*$ is unbounded. The following lemma shows that if the number of cycles is large enough compared to their individual and total weights, then there must be a subset of them of 0 total weight (i.e., a 0 skip-cycle set).

\begin{lemma}[$\star$]\label{lem:notmanyblueorredset}
Let $t \geq 3$.
Let $\mathcal{C}$ be a set of alternating cycles s.t. $|w(\mathcal{C})| \leq t$, $|w(C)| \leq 2t$ for all $ C \in \mathcal{C}$ and $|\mathcal{C}| \geq 10t^3$, then $\mathcal{C}$ contains a 0 skip-cycle set. 
\end{lemma}

Now we deal with the case when the number of cycles in $M \Delta M^*$ is bounded. In this case, for the number of edges of some color class to be unbounded, it has to be unbounded on at least one of the cycles. \Cref{lem:0skipSet1} deals with this case by first using \Cref{lem:manyzeros} to show the existence of many bundles, many 0-paths starting with a red edge and many 0-paths starting with a blue edge, or a long SAP. In the latter two cases, we can prove the existence of both positive and negative skips resulting in a 0 skip set. In the case of bundles however, we need to have both many $+1$ and many $-1$ bundles for this to work. In \Cref{lem:minusimpliesplus} we show that if the weight of an alternating cycle is bounded, then the difference between the number of $+1$ and $-1$ bundles is also bounded. This in turn allows us to prove that the existence of many bundles results in a 0 skip set as well (see \Cref{lem:notmanybundles}).


\begin{lemma}[$\star$] \label{lem:minusimpliesplus}
Let $C$ be a cycle with $|w(C)| \leq l$. If $C$ contains $3t+ l$ disjoint $-1$ (resp. $+1$) bundles, then $C$ also contains at least $t$ disjoint $+1$ (resp. $-1$) bundles. 
\end{lemma}


\begin{lemma}[$\star$]\label{lem:notmanybundles}
Let $t \geq 8\cdot 4^{\alpha}$.
Let $C$ be a cycle with $|w(C)| \leq 2t$. If $C$ contains more than $10t$ disjoint bundles then it must contain a 0 skip set. 
\end{lemma}


\begin{lemma}[$\star$]\label{lem:0skipSet1}
Let $t \geq 8\cdot4^{\alpha}$.
Let $\mathcal{C}$ be a collection of cycles s.t. $|\mathcal{C}| \leq 10t^3$, $|w(C)| \leq 2t$ for all $ C \in \mathcal{C}$ and $\mathcal{C}$ contains at least $1000t^6$ blue edges and $1000t^6$ red edges, then $\mathcal{C}$ contains a 0 skip set. 

\end{lemma}



\begin{proof}[Proof of \Cref{th:XPalpha}]
Use the algorithm of \Cref{th:falphaclose} to get a matching $M$ s.t. $k-2\cdot 4^{\alpha} \leq r(M) \leq k$.
Let $M^*$ be a PM with $k$ red edges that minimizes $|E(M \Delta M^*)|$. Consider the set of cycles $M \Delta M^*$. Observe that it cannot contain a 0 skip-cycle set (by minimality of its number of edges) and $|w(M \Delta M^*)| \leq |k-r(M)| \leq 2\cdot 4^{\alpha} $. Let $t = 256\cdot 4^{2\alpha}$ (so $t$ is large enough to apply all the previous lemmas).
If some cycle $C \in M \Delta M^*$ has $|w(C)| \geq 2t$, by \Cref{lem:0skipSet2} we get a 0 skip-cycle set. So we consider the case when all cycles $C \in M \Delta M^*$ have $|w(C)| < 2t$. If $M \Delta M^*$ contains at least $10t^3$ cycles, by \Cref{lem:notmanyblueorredset} we get a 0 skip set. So we consider the case when $|M \Delta M^*| \leq 10t^3$. By \Cref{lem:0skipSet1}, since $M \Delta M^*$ does not contain a 0 skip set, it must contain at most $f(\alpha)$ edges of some color class (for $f(\alpha) = 1000\cdot (256\cdot 4^{2\alpha})^6 = 2^{O(\alpha)}$). 
By \Cref{prop:smallsetedges} we can find a PM with exactly $k$ red edges in $n^{O(f(\alpha))}$ time if one exists.
\end{proof}

\section{Bipartite Graphs}\label{sec:XPbeta}

\begin{figure}[ht]
    \centering
    \includegraphics[scale=0.7]{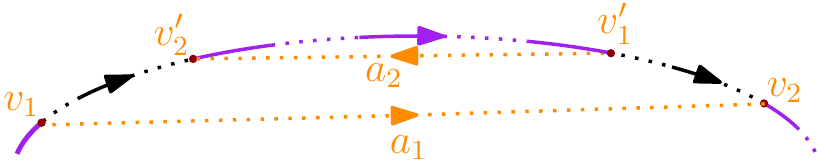}
    \caption{A biskip formed by two non-matching arcs $a_1$ and $a_2$ (in orange). Matching edges are represented by full lines and non-matching edges by dotted lines. The paths removed by the biskip are depicted in black while the rest of the alternating cycle is depicted in purple.}
    \label{fig:biskip}
\end{figure}

In this section, we consider Bipartite graphs, which contain very large independent sets ($\geq n/2$). For this reason, we instead parameterize by the bipartite independence number $\beta$.
Note that for the proof of \Cref{th:XPalpha} the only time we used the bounded independence number is in the proof of \Cref{lem:skipfrompaths}. So we need an analogue of it that works for bounded bipartite independence number, which will be given in \Cref{lem:biskipfrompaths}. We will also need a new notion of a skip that better fits the bipartite case. We call it a biskip (see \Cref{def:biskip} and \Cref{fig:biskip}). We will also rely on an orientation of the edges of the graph defined as follows. 
Given a bipartite graph $G$ with bipartition $(A,B)$ and a matching $M$, we transform $G$ into a directed graph $G_M$ by orienting every matching edge from $A$ to $B$ and every non-matching edge from $B$ to $A$.

\begin{definition}\label{def:biskip}
Let $C$ be a directed alternating cycle. A biskip $S$ is a set of 2 arcs $a_1 := (v_1,v_2)$ and $a_2 := (v_1',v_2')$ with $a_1, a_2 \notin C$ and $v_1,v_2',v_1',v_2 \in C$ (appearing in this order along $C$) s.t. $C_1 := C[v_2,v_1] \cup a_1 $ and $C_2 := C[v_2',v_1'] \cup a_2 $ are vertex disjoint directed alternating cycles, $|C| - |C_1| - |C_2| > 0$ and $|w(S)| \leq 4$ where $w(S) := w(C_1) + w(C_2) - w(C)$ is called the weight of the biskip.
\end{definition}
 
If $P \subseteq C$ is a path and $C[v_1,v_2] \subseteq P$, then we say that $P$ contains the biskip $S$. We say using $S$ to mean replacing $C$ by $C_1$ and $C_2$.
If $C \in M\Delta M'$ for some PM $M'$, then by using $S$ we also modify $M'$ accordingly (i.e. s.t. $M \Delta M'$ now contains $C_1$ and $C_2$ instead of $C$).
Two skips $\{(v_1,v_2),(v_1',v_2')\}$ and $\{(u_1,u_2),(u_1',u_2')\}$ are called disjoint if they are contained in disjoint paths.

\begin{remark*}
Note that the biskip could have been defined with one arc instead of two (since in this case one arc is enough to shorten an alternating cycle), which would have made the definition simpler. \Cref{def:biskip} is however, very similar to the definition of the skip (see \Cref{def:skip}) and this in turn allows us to prove \Cref{th:XPbeta} in an analogous way to \Cref{th:XPalpha} instead of requiring a completely different proof.
\end{remark*}

\begin{lemma}\label{lem:biskipfrompaths}
Let $P \subseteq G_M$ be a directed alternating path containing a set $\mathcal{P}$ of disjoint directed paths, each of length at least $3$ and starting and ending at a non-matching edge, s.t. $|\mathcal{P}| \geq 4^{2\beta +2}$. Then $P$ contains a biskip.
If all paths in $\mathcal{P}$ have the same weight $x$, then if $x$ is one of the following values, we get the following types of biskips:
\begin{itemize}
    \item $x=2$: negative biskip.
    \item $x=1$: negative or 0 biskip.
    \item $x=0$: positive or 0 biskip.
    \item $x=-1$: positive biskip.
\end{itemize}
\end{lemma}

\begin{proof}
Consider the graph $G_C$ defined as follows: $V(G_C) = \mathcal{P}$, there is an edge between two vertices if their corresponding paths have an arc that goes from the start vertex of the first path to the end vertex of the second.  
\begin{claim*}
$G_C$ has independence number bounded by $2\beta +1$.
\end{claim*}
\begin{claimproof}
Take any subset $Q$ of vertices of $G_C$ of size $2\beta +2$. Let $Q_1$ be $\beta + 1$ consecutive (along $C$) vertices of $Q$ and $Q_2$ the rest. Let $V_1$ be the set of start vertices of the paths corresponding to $Q_1$ in $G_M$ and $V_2$ be the set of end vertices of the paths corresponding to $Q_2$ in $G_M$. Observe that $V_1 \cup V_2$ is a balanced set of size $2\beta+2$, so there must be an arc connecting two of its vertices. Observe that the arc must be going from $V_1$ to $V_2$ since it corresponds to a non-matching edge. So $G_C$ contains an edge corresponding to this arc, i.e. $Q$ is not an independent set.
\end{claimproof}

$G_C$ must contain a clique $Q$ of size $2\beta+2$ since $|V(G_C)|= |\mathcal{P}| \geq Ram(2\beta +2, 2\beta +2)$. Let $Q_1$ be $\beta+1$ consecutive (along $C$) vertices of $Q$ and $Q_2$ the rest. Let $V_1$ be the set of end vertices of the paths corresponding to $Q_1$ in $G_M$ and $V_2$ be the set of start vertices of the paths corresponding to $Q_2$ in $G_M$. Observe that $V_1 \cup V_2$ is a balanced set of size $2\beta +2$, so there must be an arc (call it $a_1$) connecting two of its vertices. Observe that the arc must be going from $V_2$ to $V_1$ since it corresponds to a non-matching edge. Let $P_1$ and $P_2$ be the paths corresponding to its start and end vertices. $P_1$ and $P_2$ must be connected by an edge in $Q$, let $a_2$ be the corresponding arc in $G_M$. So $a_2$ connects the start of $P_1$ to the end of $P_2$ and $a_1$ connects the start of $P_2$ to the end of $P_1$. Observe that $a_1$ and $a_2$ form a biskip $S$ and $w(S) = w(a_1) + w(a_2) - w(P_1) - w(P_2)$. Let $x=w(P_1) = w(P_2)$ ($x$ depends on the type of paths considered) and note that $w(e_1),w(e_2) \in \{0,1\}$. We get $-2x \leq w(S) \leq 2-2x$ thus proving the lemma.
\end{proof}

The rest of the proof of \Cref{th:XPbeta} follows the same structure as that of \Cref{th:XPalpha} while using biskips instead of skips. Due to lack of space we will defer all the details to the appendix where we will restate all the definitions and lemmas that need to be adapted.

\section{Distance-d Independence Number}\label{sec:distD}

In this section we show that the algorithms developed for small independence number graphs cannot be generalized to distance-$d$ independence number, for $d > 2$, unless they can be used to solve EM on any graph. A distance-$d$ independent set is a set of vertices at pairwise distance
at least $d$ (i.e. the shortest path between any two of them contains at least $d$ edges) and the distance-$d$ independence number is the size of the largest such set. Note that for $d=2$ we get the normal independence number.
Let $\alpha_d(G)$ be the distance-$d$ independence number of a graph $G$. 

\begin{theorem}\label{th:dist2ind}
EM can be reduced to EM on graphs with $\alpha_d(G)=1$, for any $d > 2$, in deterministic polynomial time.

\end{theorem}
\begin{proof}
Given a graph $G = (V,E)$ we construct another graph $G' =(V', E')$ by adding two new vertices $u$ and $v$ s.t. $V' = V \cup \{u,v\}$ and $E':= E \cup (u,v) \cup \{(u,x): x \in V\}$. All edges in $E$ keep their colors while new edges get color blue. Observe that any PM on $G'$ must contain $(u,v)$ since it is the only edge connected to $v$, so by removing this edge from the PM we get a PM for $G$. Also note that $G'$ has distance-$d$ independence number of 1, for any $d > 2$, since any two vertices are connected to $u$, i.e. have distance 2. 
Now if there exists a PM $M$ with exactly $k$ red edges in $G'$, we know that $M \backslash (u,v)$ is a PM with exactly $k$ red edges in $G$.
\end{proof}

A similar theorem applies for bipartite graphs. Note that here we do not need to consider balanced independent sets (a similar result holds if we do).

\begin{theorem}\label{th:dist2indbi}
EM on bipartite graphs can be reduced to EM on bipartite graphs with $\alpha_d(G)=2$, for any $d > 2$, in deterministic polynomial time.

\end{theorem}

\begin{proof}
Given a bipartite graph $G = (U, V,E)$ we construct another bipartite graph $G' =(U', V', E')$ s.t. $U' = U \cup \{u,u'\}$ , $V' = V \cup \{v,v'\}$ and $E':= E \cup \{(u,x): x \in V'\} \cup \{(v,x): x \in U'\}$. All edges in $E$ keep their colors while new edges get color blue. Observe that any PM on $G'$ must contain $(u,v')$ and $(v,u')$ since they are the only edges connected to $u'$ and $v'$, so by removing these edges from the PM we get a PM for $G$. Also note that $G'$ has distance-$d$ independence number of 2, for any $d \geq 2$, since it can contain at most one vertex from each of $U'$ and $V'$ (any two vertices of $U'$ are connected to $v$, and any two vertices of $V'$ are connected to $u$). 
Now if there exists a PM $M$ with exactly $k$ red edges in $G'$, we know that $M \backslash ((u,v') \cup (v,u'))$ is a PM with exactly $k$ red edges in $G$.
\end{proof}

\section{Conclusion and Open Problems}\label{sec:conc}

In this paper we initiated the study of the parameterized complexity of EM by showing that it can be solved in deterministic polynomial time on graphs of bounded independence number and bipartite graphs of bounded bipartite independence number (i.e. we developed XP algorithms). This is an important step towards finding the right complexity class of the problem in general graphs as it generalizes the only previously known results on dense graph classes which were restricted to complete (bipartite) graphs. 

A natural next step would be to design corresponding FPT-algorithms in which the exponent in the running time is independent of the independence number. 
Another future direction would be to study the parameterized complexity of EM for other graph parameters. As we showed, parameterizing by higher distance independence numbers does not provide any additional structure, so it would be interesting to find other parameters that could yield non-trivial structure.
Finally, it would be interesting to prove our conjecture on the hardness of counting PMs in graphs of independence number $2$ or to find deterministic polynomial time algorithms for EM that work on graph classes for which counting PMs is \#\textbf{P}-hard.

\bibliography{references}

\begin{thebibliography}{10}

\bibitem{berger2011budgeted}
Andr{\'e} Berger, Vincenzo Bonifaci, Fabrizio Grandoni, and Guido Sch{\"a}fer.
\newblock Budgeted matching and budgeted matroid intersection via the gasoline
  puzzle.
\newblock {\em Mathematical Programming}, 128(1):355--372, 2011.

\bibitem{courcelle}
Bruno Courcelle.
\newblock The monadic second-order logic of graphs. {I}. {R}ecognizable sets of
  finite graphs.
\newblock {\em Information and Computation}, 85(1):12--75, 1990.

\bibitem{do2019party}
Norman Do.
\newblock Party problems and ramsey theory.
\newblock {\em Vinculum}, 56(2):18--19, 2019.

\bibitem{edmonds}
Jack Edmonds.
\newblock Paths, trees, and flowers.
\newblock {\em Canadian Journal of Mathematics}, 17:449--467, 1965.

\bibitem{genus}
Anna Galluccio and Martin Loebl.
\newblock On the theory of {P}faffian orientations. {I}. {P}erfect matchings
  and permanents.
\newblock {\em Electronic Journal of Combinatorics}, 6:R6, 1999.

\bibitem{geerdes}
Hans-Florian Geerdes and J{\'a}cint Szab{\'o}.
\newblock A unified proof for karzanov's exact matching theorem.
\newblock Technical Report QP-2011-02, Egerv{\'a}ry Research Group, Budapest,
  2011.
\newblock {\tt egres.elte.hu}.

\bibitem{grandoni2010optimization}
Fabrizio Grandoni and Rico Zenklusen.
\newblock Optimization with more than one budget.
\newblock {\em arXiv preprint arXiv:1002.2147}, 2010.

\bibitem{gurjar2012planarizing}
Rohit Gurjar, Arpita Korwar, Jochen Messner, Simon Straub, and Thomas Thierauf.
\newblock Planarizing gadgets for perfect matching do not exist.
\newblock In {\em International Symposium on Mathematical Foundations of
  Computer Science}, pages 478--490. Springer, 2012.

\bibitem{gurjar2017exact}
Rohit Gurjar, Arpita Korwar, Jochen Messner, and Thomas Thierauf.
\newblock Exact perfect matching in complete graphs.
\newblock {\em ACM Transactions on Computation Theory (TOCT)}, 9(2):1--20,
  2017.

\bibitem{karzanov1987maximum}
AV~Karzanov.
\newblock Maximum matching of given weight in complete and complete bipartite
  graphs.
\newblock {\em Cybernetics}, 23(1):8--13, 1987.

\bibitem{kasteleyn}
Pieter~W Kasteleyn.
\newblock Graph theory and crystal physics.
\newblock {\em Harary, F. (ed.), Graph Theory and Theoretical Physics}, pages
  43--110, 1967.

\bibitem{kruskal}
Joseph~B Kruskal.
\newblock Paths, trees, and flowers.
\newblock {\em Proceedings of the American Mathematical Society}, 7:48--50,
  1956.

\bibitem{little}
Charles H~C Little.
\newblock Kasteleyn's theorem and arbitrary graphs.
\newblock {\em Canadian Journal of Mathematics}, 25(4):758--764, 1973.

\bibitem{mastrolilli2012constrained}
Monaldo Mastrolilli and Georgios Stamoulis.
\newblock Constrained matching problems in bipartite graphs.
\newblock In {\em International Symposium on Combinatorial Optimization}, pages
  344--355. Springer, 2012.

\bibitem{mastrolilli2014bi}
Monaldo Mastrolilli and Georgios Stamoulis.
\newblock Bi-criteria and approximation algorithms for restricted matchings.
\newblock {\em Theoretical Computer Science}, 540:115--132, 2014.

\bibitem{mulmuley1987matching}
Ketan Mulmuley, Umesh~V Vazirani, and Vijay~V Vazirani.
\newblock Matching is as easy as matrix inversion.
\newblock {\em Combinatorica}, 7(1):105--113, 1987.

\bibitem{papadimitriou1982complexity}
Christos~H Papadimitriou and Mihalis Yannakakis.
\newblock The complexity of restricted spanning tree problems.
\newblock {\em Journal of the ACM (JACM)}, 29(2):285--309, 1982.

\bibitem{stamoulis2014approximation}
Georgios Stamoulis.
\newblock Approximation algorithms for bounded color matchings via convex
  decompositions.
\newblock In {\em International Symposium on Mathematical Foundations of
  Computer Science}, pages 625--636. Springer, 2014.

\bibitem{svensson2017matching}
Ola Svensson and Jakub Tarnawski.
\newblock The matching problem in general graphs is in quasi-nc.
\newblock In {\em 2017 IEEE 58th Annual Symposium on Foundations of Computer
  Science (FOCS)}, pages 696--707. Ieee, 2017.

\bibitem{yi2002matchings}
Tongnyoul Yi, Katta~G Murty, and Cosimo Spera.
\newblock Matchings in colored bipartite networks.
\newblock {\em Discrete Applied Mathematics}, 121(1-3):261--277, 2002.

\bibitem{yuster2012almost}
Raphael Yuster.
\newblock Almost exact matchings.
\newblock {\em Algorithmica}, 63(1):39--50, 2012.

\end{thebibliography}

\appendix
\section{Missing Proofs}






\begin{proof}[Proof of \Cref{lem:0skipset}]
Note that all considered skips have weight (in absolute value) in the set $\{1,2,3,4\}$. The lemma can be simply proven by enumerating all possibilities for the positive and negative skips. 

\end{proof}



\begin{proof}[Proof of \Cref{lem:manyzeros}]
Suppose (a) and (b) do not hold. We will prove (c).
By deleting a maximum size set of bundles from $P$ (i.e. only deleting their non-zero pairs not the bundle paths), we split it into at most $2t$ paths. Observe that these paths are SAPs (bundle paths only contain 0 pairs and paths between bundles cannot contain other bundles due to the maximality of the chosen set) so none of them can contain more than $t$ non-zero pairs.
By again deleting these pairs, we are left with a set of at most $2t^2$ paths $\mathcal{P}$, this time only containing 0 pairs. Note that we removed at most $2t^2$ edges from each color class, so $\mathcal{P}$ still contains at least $8t^3$ blue (resp. red) edges. 

So there must be a path $P_1 \in \mathcal{P}$ containing at least $4t$ blue (resp. red)  edges.
$P_1$ only contains 0 pairs, so it contains an equal number of matching and non-matching blue (resp. red) edges i.e. at least $2t$ matching blue (resp. red) edges. Let $\mathcal{P}_1$ be the set of paths of length 4 starting at these edges, $|\mathcal{P}_1| = 2t$. Take $\mathcal{P}'_1$
to be the set containing every second path of $\mathcal{P}_1$ (i.e. the 1st, 3rd, 5th,... paths). Observe that the paths in $\mathcal{P}'_1$ are disjoint 0-paths starting with a blue (resp. red) matching edge, and $|\mathcal{P}'_1| = t$, thus proving the lemma.

\end{proof}



\begin{proof}[Proof of \Cref{lem:manypaths}]
For each property we show how to get the desired paths:
\begin{itemize}
    \item[(a),(b)] Remove the first matching edge from the bundle paths of the $t$ bundles.
    \item[(c),(d)] Remove the first matching edge from each of the $t$ 0-paths.
    \item[(e)] For every $-1$ pair of the SAP, we take the path starting at its non-matching edge and ending at the non-matching edge of the following $+1$ pair.
    \item[(f)] For every $+1$ pair of the SAP, we take the path starting at its non-matching edge and ending at the non-matching edge of the following $-1$ pair.
\end{itemize}

\end{proof}



\begin{proof}[Proof of \Cref{lem:bigcycleskip}]
We will only prove the case $w(P) \geq 2t\cdot 4^{\alpha}$, the case $w(P) \leq -2t\cdot 4^{\alpha}$ is proven similarly. First we prove that $P$ must contain at least $t\cdot 4^{\alpha}$ disjoint $+1$ bundles. Suppose not. Let $\mathcal{B}$ be a maximum size set of disjoint $+1$ bundles in $P$. Let $P'$ be the path obtained from $P$ by contracting the bundle paths of bundles in $\mathcal{B}$. Observe that $w(P') \geq w(P) - 2(t\cdot 4^{\alpha}-1) \geq 2$. 

We claim that $P'$ cannot contain any $+1$ bundles. Suppose there exists such a bundle $B$ in $P'$. $P$ cannot contain $B$ (by maximality of $\mathcal{B}$), so $B$'s non-zero pairs are separated in $P$ by a set of $+1$ bundles $\mathcal{B}' \subseteq \mathcal{B}$ (otherwise it would still be contained in $B$). Now consider the path $P'' \subseteq P$ starting and ending with $B$'s non-zero pairs. Let $\mathcal{B}''$ be the set of $+1$ bundles formed by pairs of consecutive $+1$ pairs along $P''$. Observe that $|\mathcal{B}''| \geq |\mathcal{B}'|+1 $ and that $\mathcal{B} \backslash \mathcal{B}' \cup \mathcal{B}'' $ is a set of disjoint $+1$ bundles in $P$ of size larger than $\mathcal{B}$, a contradiction. So $P'$ cannot contain any $+1$ bundles.

Since $P'$ does not contain any consecutive $+1$ pairs, $w(P') \leq 1$ (since every $+1$ pair is now followed by a $-1$ pair), a contradiction. So $P$ must contain at least $t\cdot 4^{\alpha}$ disjoint $+1$ bundles. 
Finally we take a maximum size set of such bundles and group together every $4^{\alpha}$ consecutive ones along $P$ then use \Cref{lem:manypaths} and \Cref{lem:skipfrompaths} to get a negative skip for each group. 
\end{proof}



\begin{proof}[Proof of \Cref{lem:0skipSet2}]
Suppose $w(C) \geq 2t'$ (the case $w(C) \leq -2t'$ is treated similarly). 
By \Cref{lem:bigcycleskip}, $C$ contains at least $4t$ disjoint negative skips, of which a subset $\mathcal{S}$ of size at least $t$ must have the same weight $-w_1$, with $1 \leq w_1 \leq 4$ (by the pigeonhole principle since skips are defined to have weight $|w(S)|\leq 4$).
Now we consider 2 cases:

Case (1): $\mathcal{C}$ contains a cycle $C'$ with $w(C') \leq -t$. Then by \Cref{lem:bigcycleskip} $C'$ contains at least 4 positive skips, so $\mathcal{C}$ contains a 0 skip set by \Cref{lem:0skipset}.

Case (2): $w(C') > -t$, $\forall C' \in \mathcal{C}$. In this case $\mathcal{C}$ contains at least $4t$ negative cycles (otherwise $w(\mathcal{C}) > w(C) - 4t^2 \geq 8t^2 - 4t^2 \geq t'$), so there must be at least $4$ cycles in $\mathcal{C}$ of the same weight $-w_2$ with $1 \leq w_2 \leq t$. Observe that $w_1$ of these cycles along with $w_2$ of the skips in $\mathcal{S}$ form a 0 skip-cycle set.

\end{proof}



\begin{proof}[Proof of \Cref{lem:notmanyblueorredset}]

First note that a cycle of weight 0 is also a 0 skip-cycle set, so we will assume that no such cycle exists in $\mathcal{C}$. Now suppose $\mathcal{C}$ contains at least $4t^2$ positive and $4t^2$ negative cycles. There must be at least $2t$ cycles of same positive weight $w_1 \leq 2t$ and $2t$ cycles of same negative weight $-w_2 \geq -2t$. The set of $w_2$ cycles of weight $w_1$ and $w_1$ cycles of weight $-w_2$ is a 0 skip-cycle set. So w.l.o.g. assume $\mathcal{C}$ contains less than $4t^2$ positive cycles and let $x$ be the number of negative cycles in $\mathcal{C}$. Then $-t \leq w(\mathcal{C}) < 4t^2 \cdot 2t - x = 8t^3 -x$ so $x < 8t^3 + t < 10t^3-4t^2$. But this implies $|\mathcal{C}| < 10t^3$, a contradiction.

\end{proof}



\begin{proof}[Proof of \Cref{lem:minusimpliesplus}]
Suppose $C$ contains less than $t$ disjoint $+1$ bundles. By taking a maximum size set of disjoint $+1$ bundles and deleting their bundle paths, we are left with a set of paths $\mathcal{P}$ s.t. $|\mathcal{P}| < t$ and $w(\mathcal{P}) \geq -2t - l$. Observe that all $-1$ bundles of $C$ are still present in $\mathcal{P}$. By taking a set of disjoint $-1$ bundles of size $3t + l$ and deleting their bundle paths, we are left with a set of paths $\mathcal{P}'$ s.t. $|\mathcal{P}'| < 4t + l$ and $w(\mathcal{P}') \geq 4t + l$. 
Observe that $\mathcal{P}'$ does not contain any $+1$ bundles (by maximality of the deleted set).
So every path $P \in \mathcal{P}'$ has weight at most 1, which means $w(\mathcal{P}') < 4t + l$, a contradiction.
The other statement, with the roles of $+1$ and $-1$ bundles reversed, is proven similarly.

\end{proof}



\begin{proof}[Proof of \Cref{lem:notmanybundles}]
Suppose $C$ contains less than $2t$ disjoint $+1$ bundles.
By \Cref{lem:minusimpliesplus}, $C$ contains at most $8t$ disjoint $-1$ bundles, a contradiction to the total number of bundles. So $C$ contains at least $2t$ disjoint $+1$ bundles. Similarly, $C$ contains at least $2t$ disjoint $-1$ bundles. 

By \Cref{lem:manypaths} we get that $C$ contains at least $2t$ subpaths of weight $-1$ and $2t$ subpaths of weight $2$, all edge-disjoint, starting and ending with non-matching edges.
Now we cut $C$ into two paths $P_1$ and $P_2$, such that $P_1$ contains at least $t$ paths of weight $-1$ while $P_2$ still contains at least $t$ paths of weight $+2$ (to see that this works, simply note that by walking along the path and stopping as soon as we have covered $t$ of the subpaths of some weight, we are left with a path that contains at least $t$ subpaths of the other weight).
We divide $P_1$ (resp. $P_2$) into paths each containing at least $2\cdot4^{\alpha}$ of these subpaths (note that there are at least 4 such paths for each of $P_1$ and $P_2$). By \Cref{lem:skipfrompaths} they each contain at least one positive (resp. negative) skip.
Finally by \Cref{lem:0skipset} $C$ contains a 0 skip set.

\end{proof}



\begin{proof}[Proof of \Cref{lem:0skipSet1}]
Observe that some cycle $C_1 \in \mathcal{C}$ must contain at least $100t^3$ blue edges and some cycle $C_2 \in \mathcal{C}$ must contain at least $100t^3$ red edges. If $C_1 \neq C_2$ we let $P_1 := C_1$ and $P_2 := C_2$. Otherwise we cut $C_1$ into two paths $P_1$ and $P_2$, such that $P_1$ contains at least $50t^3$ blue edges while $P_2$ still contains at least $50t^3$ red edges (to see that this works, simply note that by walking along the path and stopping as soon as we have covered $50t^3$ of the edges of some color class, we are left with a path that contains at least $50t^3$ edges of the other class).

Now by \Cref{lem:manyzeros} we know that one of the following must be true:
\begin{itemize}
    \item[(a)] $C_1$ (resp. $C_2$) contains at least $10t$ disjoint bundles (if $C_1 \neq C_2$ then \Cref{lem:manyzeros} can be applied to both $P_1$ and $P_2$, otherwise it can be applied to $P_1 \cup P_2$).
    \item[(b)] $P_1$ (resp. $P_2$) contains an SAP with at least $5t$ non-zero pairs.
    \item[(c)] $P_1$ (resp. $P_2$) contains at least $5t$ edge-disjoint 0-paths of length at least 4 starting with a blue (resp. red) matching edge.
\end{itemize}

For case (a) we get a 0 skip set by \Cref{lem:notmanybundles}.
For cases (b) and (c), by \Cref{lem:manypaths} we get that $P_1$ (resp. $P_2$) contains at least $t$ edge-disjoint subpaths starting and ending with non-matching edges and of weight $0$ (resp. $1$). Suppose $P_1$ (resp. $P_2$) does not contain 0 skips (otherwise we are done). We divide $P_1$ (resp. $P_2$) into 4 paths each containing at least $2\cdot4^{\alpha}$ of these subpaths. By \Cref{lem:skipfrompaths} they each contain at least one positive (resp. negative) skip.
Finally by \Cref{lem:0skipset}, $\mathcal{C}$ contains a 0 skip set.

\end{proof}






\section{Bipartite Graphs of Bounded Bipartite Independence Number}

Before proving \Cref{th:XPbeta} we restate all the definitions and lemmas that need to be adapted for the use of biskips instead of skips. The proofs are analogous to the ones used for skips, we restate them for completion. 

\begin{definition}
Let $\mathcal{C}$ be a set of alternating cycles. A 0 biskip set is a set of disjoint biskips on cycles of $\mathcal{C}$ s.t. the total weight of the biskips is 0.
\end{definition}

\begin{lemma}\label{lem:0biskipset}
Let $\mathcal{S}$ be a collection of disjoint biskips. If $\mathcal{S}$ contains at least 4 positive biskips and at least 4 negative biskips (all mutually disjoint), then $\mathcal{S}$ must contain a 0 biskip set. 
\end{lemma}

\begin{proof}
Note that all considered biskips have weight (in absolute value) in the set $\{1,2,3,4\}$. The lemma can be simply proven by enumerating all possibilities for the positive and negative biskips. 
\end{proof}

\begin{lemma}\label{lem:bigcyclebiskip}
Let $P$ be an alternating path with $w(P) \geq 2t\cdot 4^{2\beta +2}$ (resp. $w(P) \leq -2t\cdot 4^{2\beta +2}$), then $P$ contains at least $t$ disjoint negative (resp. positive) biskips.
\end{lemma}

\begin{proof}
We will only prove the case $w(P) \geq 2t\cdot 4^{2\beta +2}$, the case $w(P) \leq -2t\cdot 4^{2\beta +2}$ is proven similarly. First we prove that $P$ must contain at least $t\cdot 4^{2\beta +2}$ disjoint $+1$ bundles. Suppose not. Let $\mathcal{B}$ be a maximum size set of disjoint $+1$ bundles. Let $P'$ be the path obtained from $P$ by contracting the bundle paths of bundles in $\mathcal{B}$. Observe that $w(P') \geq w(P) - 2(t\cdot 4^{2\beta +2}-1) \geq 2$. 

We claim that $P'$ cannot contain any $+1$ bundles. Suppose there exists such a bundle $B$ in $P'$. $P$ cannot contain $B$ (by maximality of $\mathcal{B}$), so $B$'s non-zero pairs are separated in $P$ by a set of $+1$ bundles $\mathcal{B}' \subseteq \mathcal{B}$ (otherwise it would still be contained in $B$). Now consider the path $P'' \subseteq P$ starting and ending with $B$'s non-zero pairs. Let $\mathcal{B}''$ be the set of $+1$ bundles formed by pairs of consecutive $+1$ pairs along $P''$. Observe that $|\mathcal{B}''| \geq |\mathcal{B}'|+1 $ and that $\mathcal{B} \backslash \mathcal{B}' \cup \mathcal{B}'' $ is a set of disjoint $+1$ bundles in $P$ of size larger than $\mathcal{B}$, a contradiction. So $P'$ cannot contain any $+1$ bundles.

Since $P'$ does not contain any consecutive $+1$ pairs, $w(P') \leq 1$ (since every $+1$ pair is now followed by a $-1$ pair), a contradiction. So $P$ must contain at least $t\cdot 4^{\alpha}$ disjoint $+1$ bundles. 
Finally we take a maximum size set of such bundles and group together every $4^{\alpha}$ consecutive ones along $C$ then use \Cref{lem:manypaths} and \Cref{lem:biskipfrompaths} to get a negative biskip for each group.  
\end{proof}

\begin{theorem} \label{th:bifalphaclose}
Given a 'Yes' instance of EM on a bipartite graph, there exists a deterministic polynomial time algorithm that outputs a PM $M$ with  $k-2\cdot 4^{2\beta +2} \leq r(M) \leq k$.
\end{theorem}

\begin{proof}
Let $M$ be a PM containing a minimum number of red edges and $M'$ a PM with a maximum number of red edges (should be at least $k$). Note that $M$ (resp. $M'$) can be computed in polynomial time by simply using a maximum weight perfect matching algorithm with $-1$ (resp. $+1$) weights assigned to red edges and $0$ weights assigned to blue edges. 

Now as long as $r(M) \leq k-2\cdot 4^{2\beta +2}$ and $r(M') > k$ we will apply the following procedure.
Let $C \in M \Delta M'$ with $w(C) > 0$ (such a cycle must exist since $r(M) < r(M')$).
If $w(C) \leq 2\cdot 4^{2\beta +2}$ then we replace $M$ by $M \Delta C$ and iterate (note that $r(M) < r(M \Delta C) \leq k$).
Otherwise, by \Cref{lem:bigcyclebiskip} $C$ contains a negative biskip. We find it in polynomial time and use it to reduce the total cycle weight, and iterate the whole procedure (note that $r(M')$ decreases). 
If at any point $r(M')$ drops below $k$, we simply output $M := M'$.
In all cases $w(M \Delta M')$ decreases after every iteration. So there can be at most $n$ iterations, each running in polynomial time.

\end{proof} 

\begin{definition}
Let $\mathcal{C}$ be a set of alternating cycles. A 0 biskip-cycle set is a set of disjoint biskips on cycles of $\mathcal{C}$ and/or cycles from $\mathcal{C}$, s.t. the total weight of the biskips minus the total weight of the cycles is 0.
\end{definition}

\begin{lemma}\label{lem:0biskipSet2}
Let $t \geq 8\cdot4^{2\beta +2}$ and $t' = 4t^2$.
Let $\mathcal{C}$ be a set of cycles and $C \in \mathcal{C}$ s.t. $|w(\mathcal{C})| \leq t'$ and $|w(C)| \geq 2t'$, then $\mathcal{C}$ contains a 0 biskip-cycle set. 

\end{lemma}

\begin{proof}
Suppose $w(C) \geq 2t'$ (the case $w(C) \leq -2t'$ is treated similarly). 
By \Cref{lem:bigcyclebiskip} $C$ contains at least $4t$ disjoint negative biskips, of which a subset $\mathcal{S}$ of size at least $t$ must have the same weight $-w_1$, with $1 \leq w_1 \leq 4$ (by the pigeonhole principle).
Now we consider 2 cases:

Case (1): $\mathcal{C}$ contains a cycle $C'$ with $w(C') \leq -t$. Then by \Cref{lem:bigcyclebiskip} $C'$ contains at least 4 positive disjoint biskips, so $\mathcal{C}$ contains a 0 biskip set by \Cref{lem:0biskipset}.

Case (2): $w(C') > -t$, $\forall C' \in \mathcal{C}$. In this case $\mathcal{C}$ contains at least $4t$ negative cycles (otherwise $w(\mathcal{C}) > w(C) - 4t^2 \geq 8t^2 - 4t^2 \geq t'$), so there must be at least $4$ cycles in $\mathcal{C}$ of the same weight $-w_2$ with $1 \leq w_2 \leq t$. Observe that $w_1$ of these cycles along with $w_2$ of the biskips in $\mathcal{S}$ form a 0 biskip-cycle set.

\end{proof}

\begin{lemma}\label{lem:binotmanybundles}
Let $t \geq 8\cdot 4^{2\beta +2}$.
Let $C$ be a cycle with $|w(C)| \leq 2t$. If $C$ contains more than $10t$ disjoint bundles then it must contain a 0 biskip set. 
\end{lemma}

\begin{proof}
Suppose $C$ contains less than $2t$ disjoint $+1$ bundles.
By \Cref{lem:minusimpliesplus}, $C$ contains at most $8t$ disjoint $-1$ bundles, a contradiction to the total number of bundles. So $C$ contains at least $2t$ disjoint $+1$ bundles. Similarly, $C$ contains at least $2t$ disjoint $-1$ bundles. 

By \Cref{lem:manypaths} we get that $C$ contains at least $2t$ subpaths of weight $-1$ and $2t$ subpaths of weight $2$, all edge-disjoint, starting and ending with non-matching edges.
Now we cut $C$ into two paths $P_1$ and $P_2$, such that $P_1$ contains at least $t$ of the $-1$-paths while $P_2$ still contains at least $t$ of the $+2$-paths (to see that this works, simply note that by walking along the path and stopping as soon as we have covered $t$ of the subpaths of some weight, we are left with a path that contains at least $t$ subpaths of the other weight).
We divide $P_1$ (resp. $P_2$) into paths each containing at least $2\cdot4^{2\beta +2}$ of these subpaths (note that there are at least 4 such paths for each of $P_1$ and $P_2$). By \Cref{lem:biskipfrompaths} they each contain at least one positive (resp. negative) biskip.
Finally by \Cref{lem:0biskipset} $C$ contains a 0 biskip set.

\end{proof}

\begin{lemma}\label{lem:0biskipSet1}
Let $t \geq 8\cdot4^{2\beta +2}$.
Let $\mathcal{C}$ be a collection of cycles s.t. $|\mathcal{C}| \leq 10t^3$, $|w(C)| \leq 2t$ for all $ C \in \mathcal{C}$ and $\mathcal{C}$ contains at least $1000t^6$ blue edges and $1000t^6$ red edges, then $\mathcal{C}$ contains a 0 biskip set. 

\end{lemma}

\begin{proof}
Observe that some cycle $C_1 \in \mathcal{C}$ must contain at least $100t^3$ blue edges and some cycle $C_2 \in \mathcal{C}$ must contain at least $100t^3$ red edges. If $C_1 \neq C_2$ we let $P_1 := C_1$ and $P_2 := C_2$. Otherwise we cut $C_1$ into two paths $P_1$ and $P_2$, such that one path contains at least $50t^3$ blue edges while $P_2$ still contains at least $50t^3$ red edges (to see that this works, simply note that by walking along the path and stopping as soon as we have covered $50t^3$ of the edges of some color class, we are left with a path that contains at least $50t^3$ edges of the other class).

Now by \Cref{lem:manyzeros} we know that one of the following must be true:
\begin{itemize}
    \item[(a)] $C_1$ (resp. $C_2$) contains at least $10t$ disjoint bundles (if $C_1 \neq C_2$ then \Cref{lem:manyzeros} can be applied to both $P_1$ and $P_2$, otherwise it can be applied to $P_1 \cup P_2$).
    \item[(b)] $P_1$ (resp. $P_2$) contains an SAP with at least $5t$ non-zero pairs.
    \item[(c)] $P_1$ (resp. $P_2$) contains at least $5t$ edge-disjoint 0-paths of length at least 4 starting with a blue (resp. red) matching edge.
\end{itemize}

For case (a) we get a 0 biskip set by \Cref{lem:binotmanybundles} (since $P_1 \cup P_2$ contains more than $10t$ bundles).
For cases (b) and (c), by \Cref{lem:manypaths} we get that $P_1$ (resp. $P_2$) contains at least $t$ edge-disjoint subpaths starting and ending with non-matching edges and of weight $0$ (resp. $1$). Suppose $P_1$ (resp. $P_2$) does not contain 0 biskips (otherwise we are done). We divide $P_1$ (resp. $P_2$) into 4 paths each containing at least $2\cdot4^{2\beta +2}$ of these subpaths. By \Cref{lem:biskipfrompaths} they each contain at least one positive (resp. negative) biskip.
Finally by \Cref{lem:0biskipset} $\mathcal{C}$ contains a 0 biskip set.

\end{proof}





\begin{proof}[Proof of \Cref{th:XPbeta}]
Use the algorithm of \Cref{th:bifalphaclose} to get a matching $M$ s.t. $k-2\cdot 4^{2\beta +2} \leq r(M) \leq k$.
Let $M^*$ be a PM with $k$ red edges that minimizes $|E(M \Delta M^*)|$. Consider the set of cycles $M \Delta M^*$. Observe that it cannot contain a 0 biskip-cycle set (by minimality of its number of edges) and $|w(M \Delta M^*)| \leq |k-r(M)| \leq 2\cdot 4^{2\beta +2} $. Let $t = 256\cdot 4^{4\beta + 4}$ (so $t$ is large enough to apply all the previous lemmas).
If some cycle $C \in M \Delta M^*$ has $|w(C)| \geq 2t$, by \Cref{lem:0biskipSet2} we get a 0 biskip-cycle set. So we consider the case when all cycles $C \in M \Delta M^*$ have $|w(C)| < 2t$. If $M \Delta M^*$ contains at least $10t^3$ cycles, by \Cref{lem:notmanyblueorredset} we get a 0 biskip set. So we consider the case when $|M \Delta M^*| \leq 10t^3$. By \Cref{lem:0biskipSet1}, since $M \Delta M^*$ does not contain a 0 biskip set, it must contain at most $f(\beta)$ matching edges of some color class (for $f(\beta) = 1000\cdot (256\cdot 4^{4\beta+4})^6 = 2^{O(\beta)}$). 
By \Cref{prop:smallsetedges} we can find a PM with exactly $k$ red edges in $n^{O(f(\beta))}$ time if one exists.

\end{proof}






\end{document}